\documentclass[pdflatex]{article}

\usepackage[pdftex]{graphicx}
\usepackage{bm}
\usepackage{braket}
\usepackage{physics}
\usepackage{amsmath}
\usepackage{amsfonts}
\usepackage{here}
\usepackage{authblk}
\usepackage{hyperref}

\begin{document}

\title{Visualization of Three-Qubit Pure States with Separation of Local and Nonlocal Degrees of Freedom}

\author{Satoru Shoji\\ email \href{mailto:satoru.shoji.t8@dc.tohoku.ac.jp}{satoru.shoji.t8@dc.tohoku.ac.jp}}

\affil{Department of Applied Physics, Graduate School of Engineering, Tohoku University, Sendai, 980-8579, Japan}

\maketitle

\abstract{Understanding the structure of multi-qubit quantum states is essential for both quantum information research and education, yet intuitive visualization beyond the single-qubit Bloch sphere remains challenging. In this work, we propose a unified geometric framework for visualizing two- and three-qubit pure states in which local degrees of freedom and entanglement degrees of freedom are explicitly separated. For two qubits, we combine Bloch-sphere representations of reduced density operators with a complex concurrence plotted on the complex plane, enabling simultaneous visualization of entanglement strength and phase structure. For three qubits, building on the generalized Schmidt decomposition, we introduce bipartite and GHZ-type tripartite complex concurrences, which, together with local Bloch vectors, provide a complete coordinate representation of the state. Unlike classification-based approaches, our method focuses on representing a given concrete state, revealing how local properties and nonlocal correlations coexist. The framework distinguishes states with identical entanglement magnitudes but different interference structures and provides intuitive insight into the balance between pairwise and genuinely tripartite entanglement. This approach offers both conceptual clarity and potential applications in quantum education and state analysis.}


\section{Introduction}

In quantum information science, understanding quantum states constitutes the foundation of theoretical research, algorithm design, and experimental analysis. However, since a quantum state is defined as a vector in a high-dimensional complex Hilbert space, its internal structure is not easily grasped intuitively. In multi-qubit systems in particular, the dimension of the state space grows exponentially, and local degrees of freedom become intricately intertwined with nonlocal correlations (entanglement). As a result, even for experts—let alone beginners—it is often difficult to develop an intuitive understanding of the structure of a concrete quantum state.

In the case of a single qubit, any pure and mixed states can be uniquely represented as a point in the Bloch sphere. This geometric representation has provided a powerful pedagogical framework for understanding quantum states, unitary operations, and measurement processes in an intuitive manner. Indeed, the Bloch sphere may be regarded as one of the most successful visualization tools in quantum mechanics education.

In recent years, significant efforts have been devoted to extending such geometric intuition to multi-qubit systems. For example, Ref.~\cite{chang2024geometric} systematically organizes the geometric features of single-qubit and entangled states and demonstrates the potential of visualization-based approaches. Ref.~\cite{bley2024visualizing} proposes a framework for representing entanglement in multi-qubit systems as geometric structures, discussing the correspondence between invariants and visual representations. Ref.~\cite{ruan2023venus} introduces a method for visualizing quantum states as novel geometric objects, enabling intuitive understanding of superposition and entanglement. Ref.~\cite{barthe2023bloch} presents a hierarchical Bloch-sphere representation of multi-qubit pure states, visualizing state structure in a tree-like manner. 

From an educational perspective, visualization has also been reported to enhance conceptual understanding. A pilot study on quantum information education using visualization tools~\cite{bley2025visualizing} suggests that visual aids significantly improve problem-solving performance and conceptual comprehension compared to purely symbolic manipulation. Moreover, research on interactive teaching materials and digital simulations continues to advance~\cite{zaman2022student, kohnle2014investigating, qerimi2025comparing}.

These studies clearly demonstrate that geometric understanding of multi-qubit states is important both theoretically and pedagogically. Nevertheless, many existing approaches primarily focus on entanglement classification, geometric interpretations of invariants, or global structural analysis of the state space. The perspective of representing an arbitrary concrete quantum state as coordinates that explicitly separate local and nonlocal degrees of freedom has not yet been fully systematized.

From an educational standpoint, the following features are particularly important:
\begin{enumerate}
    \item A clear distinction between degrees of freedom that change under local unitary transformations and those that remain invariant,
    \item A separation between the \emph{magnitude} of entanglement and its \emph{phase structure},
    \item Visualization of the decomposition into bipartite and tripartite correlations.
\end{enumerate}
A representation that simultaneously satisfies these requirements would provide a powerful pedagogical foundation for structural understanding of multi-qubit systems.

The aim of this work is to construct a visualization framework for two- and three-qubit pure states in which local degrees of freedom and entanglement degrees of freedom are explicitly separated. For the two-qubit case, we employ the Schmidt decomposition as a foundation. The reduced density operator of each qubit is represented on the Bloch sphere, while entanglement is expressed as a complex concurrence, which is defined in this study, plotted on the complex plane. This representation enables simultaneous visualization of the strength of entanglement (its modulus) and the phase information (its argument).

For the three-qubit case, we extend this framework based on the generalized Schmidt decomposition~\cite{GSD, acin2001three}. Bipartite concurrences and GHZ-type tripartite concurrence are generalized into complex concurrences and displayed on the complex plane. At the same time, the local state of each qubit is represented on the Bloch sphere. In this way, general states in which bipartite and tripartite correlations coexist can be expressed as unified geometric coordinates.

A distinctive feature of our approach is that it does not aim at classification of states. Instead, it provides a coordinate representation that explicitly reveals how a given concrete state is composed of local and nonlocal degrees of freedom. As a consequence,
\begin{itemize}
    \item states with identical entanglement magnitude but different interference structures can be distinguished,
    \item the distinction between GHZ-type and W-type states becomes visually transparent,
    \item the coexistence of bipartite and tripartite correlations can be intuitively understood.
\end{itemize}

The remainder of this paper is organized as follows. In Sec.~\ref{sec:twoQviz}, we describe the decomposition of degrees of freedom and visualization scheme for two-qubit pure states. In Sec.~\ref{sec:threeQviz}, we extend the framework to three-qubit systems. Finally, Sec.~\ref{sec:conclusion} summarizes the main features of the method and discusses future directions.

\section{Visualization of Two-Qubit States}\label{sec:twoQviz}

In this section, we decompose the degrees of freedom of a two-qubit pure state into local and entanglement contributions, and present a method to visualize them in a unified framework.

\subsection{Schmidt Decomposition and Separation of Degrees of Freedom}

For an arbitrary two-qubit pure state $\ket{\psi_{12}}$ defined on the composite system of qubit $1$ and qubit $2$, the Schmidt decomposition takes the form
\begin{align}
    \ket{\psi_{12}} = (U_1 \otimes U_2) (\lambda_0\ket{00} + \lambda_1\ket{11}), 
    \quad 0 \le \lambda_1 \le \lambda_0, \quad \lambda_0^2 + \lambda_1^2 = 1,
    \label{eq:2q_shumidt_1}
\end{align}
where $U_1$ and $U_2$ are local special unitary operators acting on qubits $1$ and $2$, respectively, and $\lambda_0, \lambda_1$ are real Schmidt coefficients.

The decomposition in Eq.~\eqref{eq:2q_shumidt_1} is not unique. Because any single-qubit special unitary can be decomposed as
\[
U_j = R_z(\phi_j) R_y(\theta_j) R_z(\phi'_j), \quad \phi_j, \theta_j, \phi_j^{'} \in \mathbb{R} \quad j \in \{1,2\},
\]
with $R_z(\phi) = \mathrm{e}^{-i Z \phi/2}$ and $R_y(\theta) = \mathrm{e}^{-i Y \theta/2}$, the following two expressions
\begin{align}
    &\left( R_z(\phi_1) R_y(\theta_1) R_z(\phi') 
    \otimes R_z(\phi_2) R_y(\theta_2) R_z(0) \right)
    (\lambda_0\ket{00} + \lambda_1\ket{11}), \\
    &\left( R_z(\phi_1) R_y(\theta_1) R_z(0) 
    \otimes R_z(\phi_2) R_y(\theta_2) R_z(\phi') \right)
    (\lambda_0\ket{00} + \lambda_1\ket{11})
\end{align}
represent the same state,
\begin{equation}
    \left( R_z(\phi_1) R_y(\theta_1) 
    \otimes R_z(\phi_2) R_y(\theta_2) \right)
    \left( \mathrm{e}^{-i\phi'/2}\lambda_0\ket{00} 
    + \mathrm{e}^{i\phi'/2}\lambda_1\ket{11} \right).
\end{equation}

Thus, one of the three degrees of freedom of $\mathrm{SU}(2)$ can be absorbed into the relative phase between $\ket{00}$ and $\ket{11}$.  
To achieve a clear separation between local and entanglement degrees of freedom, we adopt the following decomposition:
\begin{align}
    \ket{\psi_{12}} 
    = (\tilde{U}_1 \otimes \tilde{U}_2)
    (\lambda_0\ket{00} + \mathrm{e}^{i\alpha}\lambda_1\ket{11}),
    \quad \alpha \in [0, 2\pi).
    \label{eq:2q_shumidt_2}
\end{align}
Here,
\begin{equation}
    \tilde{U}_j = R_z(\phi_j) R_y(\theta_j),
    \quad j \in \{1,2\}.
    \label{eq:U_tilde_def}
\end{equation}

A two-qubit pure state is specified by four complex amplitudes subject to normalization. Accounting for the irrelevance of global phase, the total number of independent real parameters is $2\times4 - 2 = 6$.  
In Eq.~\eqref{eq:2q_shumidt_2}, these six degrees of freedom are distributed as follows: four parameters in $\tilde{U}_1$ and $\tilde{U}_2$ (two each), one parameter in $\lambda_0$, and one in $\alpha$.

The parameters $\phi_j$ and $\theta_j$ correspond to the longitude and latitude, respectively, of the reduced density operators
\begin{align}
    \rho_1 &= \Tr_2(\ketbra{\psi_{12}}{\psi_{12}}), \\
    \rho_2 &= \Tr_1(\ketbra{\psi_{12}}{\psi_{12}})
\end{align}
when represented on the Bloch sphere.

\subsection{Concurrence and Complex Concurrence}

The concurrence $C$ of a two-qubit pure state $\ket{\psi_{12}}$~\cite{hill1997entanglement} can be computed from the Schmidt coefficients as
\begin{equation}
    C = 2\lambda_0 \lambda_1.
\end{equation}
Concurrence quantifies the strength of entanglement: $C=0$ corresponds to a separable state, while $C=1$ indicates maximal entanglement.

We introduce the \emph{complex concurrence} $\tilde{C}$ defined by
\begin{equation}
    \tilde{C} = 2 \mathrm{e}^{i\alpha} \lambda_0 \lambda_1.
    \label{eq:complex_conc_def_2}
\end{equation}
The concurrence is the modulus of the complex concurrence. While $C$ is invariant under local unitaries, the argument of $\tilde{C}$ changes under local unitary transformations.

For example, consider the Bell family
\begin{align}
    \ket{\Phi(\alpha)} = \frac{1}{\sqrt{2}}(\ket{00} + \mathrm{e}^{i\alpha}\ket{11}).
\end{align}
The concurrence is always $C = |\tilde{C}| = 1$, indicating identical entanglement strength. However, the argument of $\tilde{C}$ equals $\alpha$, directly reflecting the relative phase between $\ket{00}$ and $\ket{11}$. This phase can be modified by local unitaries.

For a non-maximally entangled state
\begin{equation}
    \lambda_0\ket{00} + \mathrm{e}^{i\alpha}\lambda_1\ket{11},
\end{equation}
the magnitude $C = |\tilde{C}| = |2\mathrm{e}^{i\alpha}\lambda_0\lambda_1| = 2\lambda_0\lambda_1$ captures only the entanglement strength, whereas the phase $\alpha$ is lost in $C$. In contrast, the argument of $\tilde{C}$ retains this phase information and characterizes a two-body interference structure that does not appear in the local reduced density operators
\begin{align}
    \rho_j = 
    \begin{pmatrix}
        \lambda_0^2 & 0 \\
        0 & \lambda_1^2
    \end{pmatrix}.
\end{align}

\subsection{Visualization}
\label{sec:two_Q_viz}

\begin{figure}[H]
    \centering
    \includegraphics[width=0.8\linewidth]{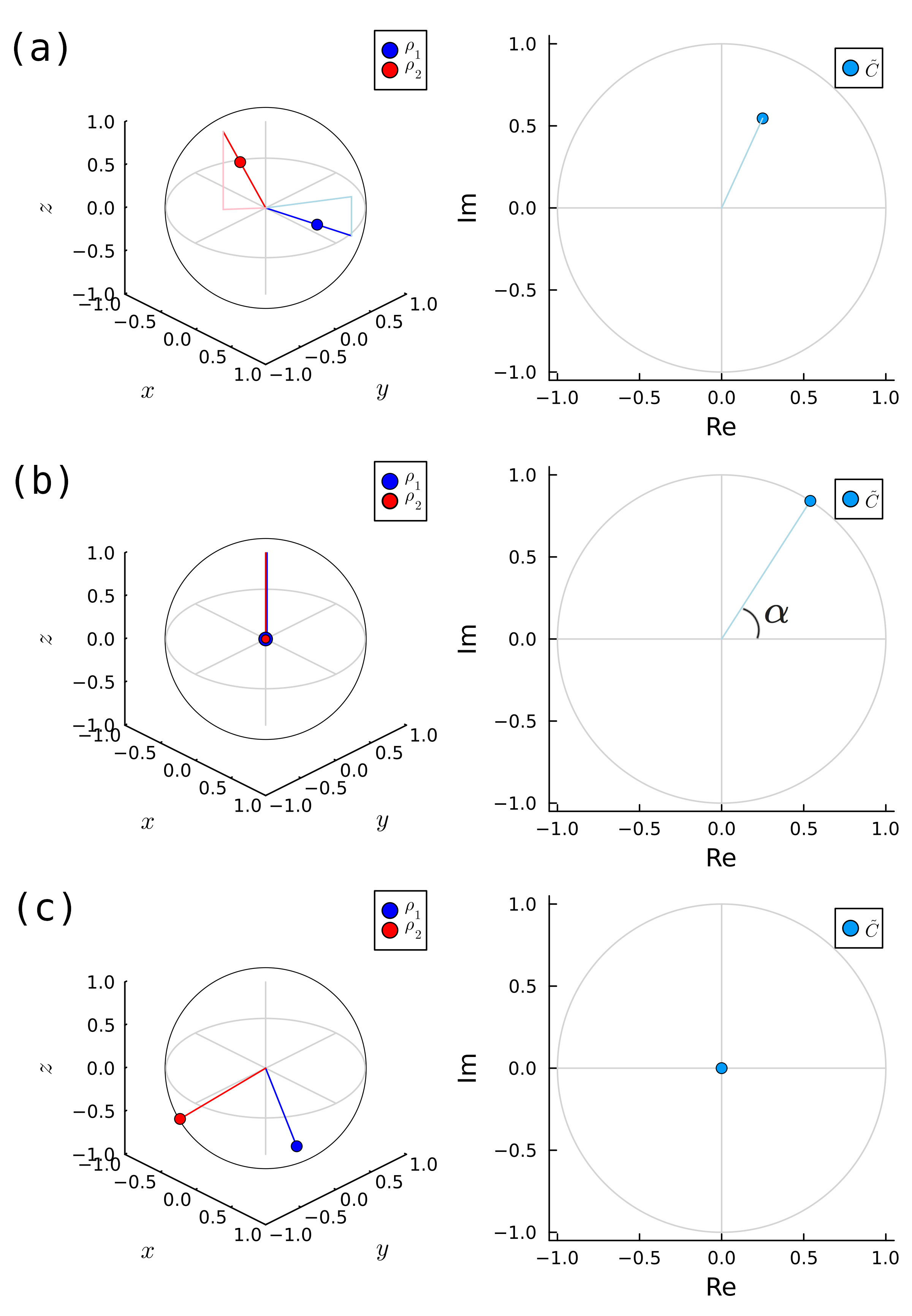}
    \caption{
    Examples of two-qubit state visualization in the proposed framework.
    (a) A general two-qubit state. The left panel shows the reduced density operators $\rho_1$ and $\rho_2$ in Bloch sphere, while the right panel shows the complex concurrence $\tilde{C}$ on the complex plane.
    (b) Visualization of $(\ket{00} + \mathrm{e}^{i\alpha} \ket{11})/\sqrt{2}$. Both reduced density operators are located at the origin (maximally mixed states). The complex concurrence lies on the unit circle at $\mathrm{e}^{i\alpha}$.
    (c) Example of a product state. The local states are pure and thus lie on the surface of the Bloch sphere. Since the state is separable, the complex concurrence is zero.
    }
    \label{fig:2q}
\end{figure}

Fig.~\ref{fig:2q} illustrates examples of the visualization scheme for two-qubit states.An arbitrary $\ket{\psi_{12}}$ can be represented by the complex concurrence together with the longitude and latitude $(\phi_1,\theta_1)$ and $(\phi_2,\theta_2)$ of the reduced density operators $\rho_1$ and $\rho_2$.  
By representing $\tilde{C}$ on the complex plane and $\rho_1$, $\rho_2$ on Bloch spheres, we obtain a complete visualization of the two-qubit pure state.  
This framework provides a clear separation between local and nonlocal degrees of freedom.

When the state is maximally entangled, the visualization is not unique due to the ricochet property:
\begin{align}
    \frac{1}{\sqrt{2}}
    (R_z(\phi)R_y(\theta) \otimes I)
    (\ket{00} + \mathrm{e}^{i\alpha}\ket{11})
    =
    \frac{1}{\sqrt{2}}
    (I \otimes R_z(\alpha)R_y(-\theta))
    (\ket{00} + \mathrm{e}^{i\phi}\ket{11}).
\end{align}
In such cases, we fix $(\phi_2,\theta_2) = (0,0)$ and adopt the canonical representation
\begin{align}
    \frac{1}{\sqrt{2}}
    (R_z(\phi_1)R_y(\theta_1) \otimes I)
    (\ket{00} + \mathrm{e}^{i\alpha}\ket{11}).
\end{align}
However, when visualizing dynamics, this convention may be relaxed in order to maintain continuity between successive states.

\section{Visualization of Three-Qubit States}\label{sec:threeQviz}

In this section, we extend the visualization scheme introduced for two-qubit states to the case of three-qubit pure states.

\subsection{Generalized Schmidt Decomposition and Separation of Local Degrees of Freedom}

For a three-qubit pure state $\ket{\psi_{123}}$, a generalized Schmidt decomposition (GSD) that is invariant under local unitaries has been proposed~\cite{GSD, acin2001three}:
\begin{align}
    &\ket{\psi_{123}} = (U_1 \otimes U_2 \otimes U_3)
    (\lambda_0\ket{000} + \mathrm{e}^{i\varphi} \lambda_1\ket{100} + \lambda_2\ket{101} + \lambda_3\ket{110} + \lambda_4\ket{111}),
    \label{eq:3q_gsd_1}\nonumber\\
    &\lambda_j\ge0, \quad \sum_{j=0}^4 \lambda_j^2 = 1, \quad 0\le\phi\le\pi.
\end{align}
Here, $U_j~(j=1,2,3)$ are local special unitary operators acting on the $j$th qubit.

As in the two-qubit case, in order to separate local and entanglement degrees of freedom, we rewrite Eq.~\eqref{eq:3q_gsd_1} in a form analogous to Eq.~\eqref{eq:2q_shumidt_2}:
\begin{align}
    \ket{\psi_{123}} = (\tilde{U}_1 \otimes \tilde{U}_2 \otimes \tilde{U}_3)
    (&\lambda_0\ket{000} + \mathrm{e}^{i\alpha_1} \lambda_1\ket{100} + \mathrm{e}^{i\alpha_2} \lambda_2\ket{101}
    \nonumber\\
    &+ \mathrm{e}^{i\alpha_3} \lambda_3\ket{110} + \mathrm{e}^{i\alpha_4} \lambda_4\ket{111}),
    \label{eq:3q_gsd_2}
\end{align}
where $\tilde{U}_j$ are defined, similarly to Eq.~\eqref{eq:U_tilde_def}, by
\begin{equation}
    \tilde{U}_j = R_z(\phi_j) R_y(\theta_j), \quad \phi_j,\theta_j \in \mathbb{R}, \quad j \in \{1,2,3\}.
    \label{eq:U_tilde_def3}
\end{equation}

If we decompose the local unitaries in Eq.~\eqref{eq:3q_gsd_1} as $U_j = R_z(\phi_j) R_y(\theta_j) R_z(\phi'_j)$, then the phase $\varphi$ in Eq.~\eqref{eq:3q_gsd_1} and the phases $\alpha_j$ in Eq.~\eqref{eq:3q_gsd_2} are related by
\begin{align}
    \alpha_1 &= \varphi + \phi_1',\\
    \alpha_2 &= \phi_1' + \phi_3',\\
    \alpha_3 &= \phi_1' + \phi_2',\\
    \alpha_4 &= \phi_1' + \phi_2' + \phi_3'.
\end{align}
Here, we have fixed the global phase such that the coefficient of $\ket{000}$ is real. Without loss of generality, we choose $\alpha_j \in [0, 2\pi)$.

\subsection{Two-Body and Three-Body Concurrences and Complex Concurrences}\label{sec:three_conc}

Based on the decomposition in Eq.~\eqref{eq:3q_gsd_2}, the concurrences can be computed as follows~\cite{hill1997entanglement}:
\begin{align}
    C_{12} &= 2 \lambda_0 \lambda_3,
    \label{eq:3q_conc_def_start}\\
    C_{13} &= 2 \lambda_0 \lambda_2,\\
    C_{23} &= 2 \left| \mathrm{e}^{i(-\alpha_1+\alpha_4)} \lambda_1 \lambda_4 - \mathrm{e}^{i(-\alpha_2+\alpha_3)} \lambda_2 \lambda_3 \right|,\\
    C_{123} &= 2 \lambda_0 \lambda_4.
    \label{eq:3q_conc_def_end}
\end{align}
Here, $C_{jk}$ denotes the concurrence between the $j$th and $k$th qubits, while $C_{123}$ is a GHZ-type (three-body) concurrence for the entire system.

We briefly clarify the relation between the concurrences defined in Eqs.~\eqref{eq:3q_conc_def_start}--\eqref{eq:3q_conc_def_end} and standard notions of three-qubit entanglement.
The quantities $C_{12}$, $C_{13}$, and $C_{23}$ are the usual bipartite concurrences defined for the corresponding two-qubit reduced subsystems, and they quantify the strength of pairwise entanglement present in a three-qubit state.
In contrast, $C_{123}$ can be nonzero even when all two-qubit reduced subsystems are separable.
It attains the maximum value $1$ for GHZ-type tripartite entanglement and vanishes for the W state.
Using the 3-tangle $\tau_{123}$ introduced by Coffman, Kundu, and Wootters~\cite{distr_ent}, one can write
\begin{equation}
    C_{123} = \sqrt{\tau_{123}}.
\end{equation}
W-type entanglement is characterized by vanishing $C_{123}$ but nonzero pairwise concurrences.
Thus, the four concurrences in Eqs.~\eqref{eq:3q_conc_def_start}--\eqref{eq:3q_conc_def_end} explicitly provide the decomposition of correlations into pairwise (two-body) contributions and a GHZ-type (three-body) contribution. Treating them jointly enables a systematic characterization of entanglement structure in a general three-qubit pure state.

Analogously to the two-qubit case (Eq.~\eqref{eq:complex_conc_def_2}), we define the corresponding \emph{complex concurrences}:
\begin{align}
    \tilde{C}_{12} &= 2 \mathrm{e}^{i\alpha_3} \lambda_0 \lambda_3,
    \label{eq:complex_conc_def_3_st}\\
    \tilde{C}_{13} &= 2 \mathrm{e}^{i\alpha_2} \lambda_0 \lambda_2,\\
    \tilde{C}_{23} &= 2 \mathrm{e}^{-2i\alpha_1} \left( \mathrm{e}^{i(\alpha_1+\alpha_4)} \lambda_1 \lambda_4 - \mathrm{e}^{i(\alpha_2+\alpha_3)} \lambda_2 \lambda_3 \right),\\
    \tilde{C}_{123} &= 2 \mathrm{e}^{i\alpha_4} \lambda_0 \lambda_4.
    \label{eq:complex_conc_def_3_end}
\end{align}
The concurrence is the modulus of the complex concurrence and is invariant under local unitaries.
Moreover, $\tilde{C}_{12}$ is invariant under local unitaries acting on the third qubit, and analogous statements hold for $\tilde{C}_{13}$ and $\tilde{C}_{23}$.

Given $\tilde{C}_{12}$, $\tilde{C}_{13}$, $\tilde{C}_{23}$, and $\tilde{C}_{123}$, Eqs.~\eqref{eq:complex_conc_def_3_st}--\eqref{eq:complex_conc_def_3_end} determine the nonlocal part of Eq.~\eqref{eq:3q_gsd_2}, i.e., the component independent of the local unitaries.
Therefore, within our framework, these four complex concurrences encode all entanglement information of $\ket{\psi_{123}}$.
In addition, if the single-qubit reduced density matrices $\rho_j$ are known, then the parameters $(\phi_j,\theta_j)$ specifying the local unitaries can be determined, and the state $\ket{\psi_{123}}$ is fully specified.

A three-qubit pure state has $14$ real degrees of freedom after removing the global phase, and these are distributed as
\begin{itemize}
    \item the complex concurrences $\tilde{C}_{12}, \tilde{C}_{13}, \tilde{C}_{23}, \tilde{C}_{123}$,
    \item the local-state parameters $(\phi_1,\theta_1)$, $(\phi_2,\theta_2)$, $(\phi_3,\theta_3)$.
\end{itemize}

\subsection{Visualization}

\begin{figure}[H]
    \centering
    \includegraphics[width=0.7\linewidth]{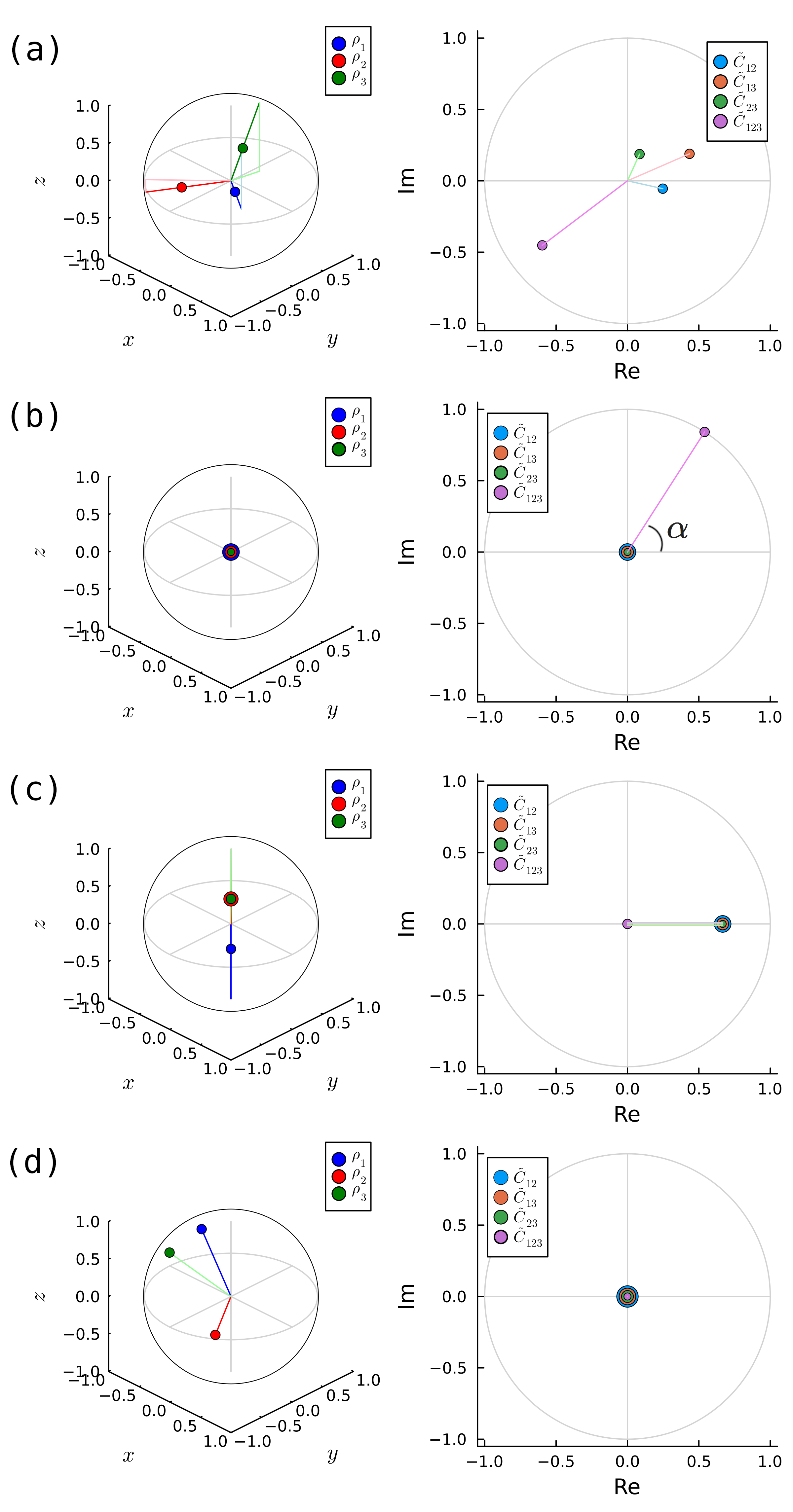}
    \caption{Examples of three-qubit state visualization in the proposed framework.
    (a) A general three-qubit state. The left panel shows three reduced density operators $\rho_1, \rho_2, \rho_3$ on Bloch spheres, and the right panel shows the four complex concurrences $\tilde{C}_{12}, \tilde{C}_{13}, \tilde{C}_{23}, \tilde{C}_{123}$ on the complex plane.
    (b) Visualization of $(\ket{000} + \mathrm{e}^{i\alpha} \ket{111})/\sqrt{2}$. All reduced density operators lie at the origin (maximally mixed states). Only the GHZ-type complex concurrence $\tilde{C}_{123}$ lies on the unit circle at $\mathrm{e}^{i\alpha}$, while the others vanish.
    (c) Visualization of $(\ket{000}+\ket{101}+\ket{110})/\sqrt{3}$. The GHZ-type complex concurrence $\tilde{C}_{123}$ is zero, and the bipartite complex concurrences satisfy $\tilde{C}_{12}=\tilde{C}_{13}=\tilde{C}_{23}=2/3$.
    (d) Example of a product state. Since the local states are pure, the points lie on the surfaces of the Bloch spheres. As the state is separable, all complex concurrences shown in the right panel are zero.
    }
    \label{fig:3q}
\end{figure}

Fig.~\ref{fig:3q} shows examples of the proposed visualization of three-qubit states.
The local states are displayed on Bloch spheres via the reduced density matrices, and the complex concurrences are displayed on the complex plane, yielding an intuitive representation of a three-qubit pure state.

In our representation, each qubit's local state is visualized on the Bloch sphere through its reduced density matrix, while the bipartite entanglement and GHZ-type tripartite entanglement are visualized on the complex plane as the complex concurrences $\tilde{C}_{12}$, $\tilde{C}_{13}$, $\tilde{C}_{23}$, and $\tilde{C}_{123}$.
Here, the moduli of the complex concurrences represent the entanglement strengths, whereas the arguments encode phase information arising from interference structure; this phase information is genuinely nonlocal and does not appear directly in local reduced density operators.
Consequently, even for states with identical entanglement magnitudes, our method enables an intuitive discrimination of differences in the balance between pairwise and tripartite correlations, as well as differences in phase structure.
In particular, a notable feature of our approach is that it provides a unified visualization framework for general three-qubit pure states in which bipartite entanglement and GHZ-type tripartite entanglement can coexist.

We emphasize that the visualization in this framework is not unique. The non-uniqueness arises the properties of maximally entangled states. The non-uniqueness is of the same origin as the ricochet property discussed in Sec.~\ref{sec:two_Q_viz}. Since the visualization of maximally entangled two-qubit states is not unique in our framework, the visualization of three-qubit states that can be written as the tensor product of a maximally entangled two-qubit state and a single-qubit state is also not unique.

\section{Conclusion and Outlook}\label{sec:conclusion}

In this work, we proposed a visualization framework for two- and three-qubit pure states that explicitly separates local degrees of freedom from entanglement degrees of freedom. For two qubits, we showed that plotting the reduced density operators on Bloch spheres and representing entanglement on the complex plane via the complex concurrence enables one to visualize both the entanglement strength and the associated phase structure simultaneously. For three qubits, building on the generalized Schmidt decomposition, we introduced bipartite and GHZ-type tripartite complex concurrences and demonstrated that, together with the local reduced states, they provide a unified coordinate representation that resolves the contributions of pairwise and genuinely tripartite correlations. A central feature of our approach is that it is not aimed at classifying states; rather, it geometrically exposes the structure of a given concrete state by decomposing it into local and nonlocal components. This perspective offers an intuitive means to distinguish, for example, different interference structures among states sharing the same entanglement magnitude, as well as qualitative differences between GHZ-type and W-type entanglement.

An important direction for future work is to extend the framework to four or more qubits. In four-qubit systems, genuinely four-partite entanglement emerges in addition to two- and three-body correlations, and the number of degrees of freedom increases substantially. One possible route is to generalize higher-order invariants, such as the four-tangle, to complex-valued quantities and to visualize them in a hierarchical manner, thereby capturing the structure of multipartite entanglement beyond the three-qubit regime. Further challenges include extending the method to mixed states and visualizing dynamical processes such as unitary time evolution and noise channels induced by quantum circuits. An interactive implementation that allows users to track state changes under circuit operations would be valuable for both educational and research purposes, providing an intuitive platform for exploring the geometry and dynamics of entanglement.

\section*{Acknowledgments}
This research did not receive any specific grant from funding agencies in the public, commercial, or not-for-profit sectors.

\bibliographystyle{unsrt}
\bibliography{bibs}

\end{document}